\documentclass[aps,superscriptaddress,amsmath,amssymb,floatfix]{PoS}

\usepackage{epsfig}
\usepackage{amssymb}

\newcommand{\agev}{\mbox{$A$~GeV}}               
\newcommand{\gevc}{\mbox{GeV$/c$}}
\newcommand{\gevcc}{\mbox{GeV$/c^2$}}

\newcommand{\rb}[1]{\mbox{\textrm{\scriptsize #1}}}

\newcommand{\pt}{\ensuremath{p_{\rb{T}}}}
\newcommand{\mt}{\ensuremath{m_{\rb{T}}}}
\newcommand{\vi}{\ensuremath{v_{\rb{2}}}}
\newcommand{\phiep}{\ensuremath{\Phi_{\rb{2EP}}}}
\newcommand{\philab}{\ensuremath{\phi_{\rb{lab}}}}
\newcommand{\phir}{\ensuremath{\Phi_{\rb{r}}}}

\title{Anisotropic Flow of Strange Particles at SPS}

\ShortTitle{Anisotropic Flow of Strange Particles at SPS}

\author{\speaker{Grzegorz~Stefanek} for the NA49 Collaboration\\
        Swietokrzyska Academy, Kielce, Poland\\
        E-mail: \email{stefanek@pu.kielce.pl}

\vspace{0.5cm} \noindent C.~Alt$^{9}$, T.~Anticic$^{23}$,
B.~Baatar$^{8}$,D.~Barna$^{4}$, J.~Bartke$^{6}$, L.~Betev$^{10}$,
H.~Bia{\l}\-kowska$^{20}$, C.~Blume$^{9}$,  B.~Boimska$^{20}$,
M.~Botje$^{1}$
% ...

}

\author{\\
\noindent
$^{1}$NIKHEF, Amsterdam, Netherlands. \\
% ...
$^{4}$KFKI Research Institute for Particle and Nuclear Physics, Budapest, Hungary.\\
% ...
$^{6}$Institute of Nuclear Physics, Cracow, Poland.\\
% ...
$^{8}$Joint Institute for Nuclear Research, Dubna, Russia.\\
$^{9}$Fachbereich Physik der Universit\"{a}t, Frankfurt, Germany.\\
$^{10}$CERN, Geneva, Switzerland.\\
% ...
$^{20}$Institute for Nuclear Studies, Warsaw, Poland.\\
% ...
$^{23}$Rudjer Boskovic Institute, Zagreb, Croatia.\\

}

 \abstract{

 The elliptic flow for Lambda hyperons and $K^{0}_{s}$ mesons was measured
by the NA49 experiment in semicentral Pb+Pb collisions at 158A
GeV. The standard method of correlating particles with an event
plane has been used. Measurements of \vi~ near mid-rapidity are
reported as a function of centrality, rapidity and transverse
momentum. Elliptic flow of $\Lambda$ and $K^{0}_{s}$ particles
increases both with the impact parameter and with the transverse
momentum. It is compared with \vi~ for pions and protons as well
as with various model predictions. The NA49 results are compared
with data from NA45/CERES and STAR experiments. }

\FullConference{The 3rd edition of the International Workshop --- The Critical Point and Onset of Deconfinement --- \\
         July 3-7 2006\\
         Galileo Galilei Institute, Florence, Italy}

\begin{document}

\section{Introduction}

Elliptic flow has its origin in the spatial anisotropy of the
initial reaction volume in non-central collisions and in particle
rescatterings in the evolving system which convert the spatial
anisotropy into a momentum anisotropy \cite{Ollitrault:1997vz}.
The spatial anisotropy decreases rapidly because of the fast
expansion of the system \cite{Kolb:2000sd} making the momentum
anisotropy measured at the end of this evolution strongly
dependent on the matter properties and the effective equation of
state (EoS) at the early stage \cite{early_stages,Teaney:2000cw}.
It is particularly sensitive to the degree of thermalization in
the produced particle system. Flow of heavy particles is affected
more strongly by changes in the EoS than flow of pions
\cite{Teaney:2000cw,Huovinen:2001cy,Snellings:2003mh}. Moreover,
various hadron types are believed to decouple at different times
and with different temperatures \cite{vanHecke:1998yu}. Thus the
elliptic flow of various particle species allows insight into
different stages of the collision. Comparison of measured
anisotropies of different particle species with various model
calculations, for example hydrodynamical or quark coalescence,
provides an important test of various evolution scenarios.

The anisotropic flow parameters measured to date at SPS and lower
energies are mainly those of pions and protons
\cite{v2_pi_p,Alt:2003ab}. We have extended elliptic flow
measurements in 158\agev~Pb+Pb ($\sqrt{s_{\rb{NN}}}=17.2$ GeV)
collisions to $\Lambda$ hyperons and $K^{0}_{s}$ particles to test
the validity of the hydrodynamic scenario and check the degree of
thermalization at SPS energies.

\section{Experiment and Data}

The NA49 experimental setup is shown in Fig.~1. The main
components
\begin{figure}
\begin{center}
\includegraphics[width=0.8\textwidth]{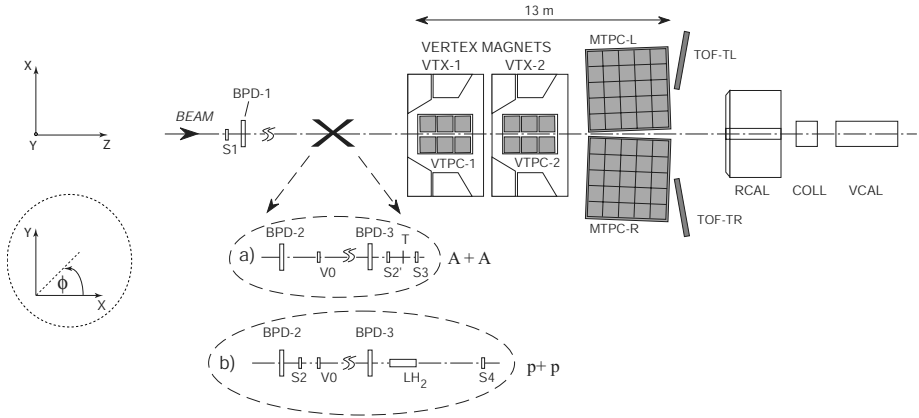}
\caption{The experimental setup of the NA49 experiment}
\end{center}
\label{fig1}
\end{figure}
of the NA49 detector \cite{NA49_setup} are four large-volume Time
Projection Chambers for tracking and particle identification. The
TPC system consists of two vertex chambers inside the spectrometer
magnets which allow separation of positively and negatively
charged tracks and a precise measurement of the particle momenta
with a resolution $\sigma(p)/p^{2}$ =
$(0.3-7)\times10^{-4}(\gevc)^{-1}$. Two main chambers, placed
behind the magnets at both sides of the beam, were optimized for
high precision detection of the ionization loss $dE$/$dx$ with a
resolution of 3$-$6\%. Downstream of the TPCs a veto calorimeter
detects projectile spectators and is used for triggering and
centrality selection. The data sample consists of $3\times10^{6}$
semi-central Pb+Pb events after online trigger selection of the
23.5\% most central collisions. The events were divided into three
different centrality bins, which correspond to the first three
bins used in a previous analysis (see Table 1 in
\cite{Alt:2003ab}). They are defined by centrality ranges 0-5\%
(bin 1), 5-12.5\% (bin 2), and 12.5-23.5\% (bin 3) which
correspond to impact parameter ranges which are: 0-3.4 fm (bin 1),
3.4-5.3 fm (bin 2), and 5.3-7.4 fm (bin 3).
\begin{figure}
\begin{center}
\includegraphics[width=0.7\textwidth]{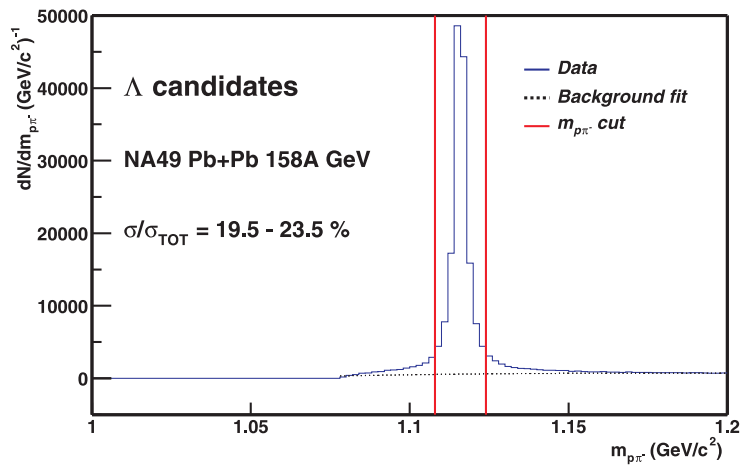}
\caption{The invariant mass distribution of p$\pi^{-}$ candidate
pairs in Pb+Pb collisions in the centrality range
$\sigma$/$\sigma_{TOT}$ = 19.5$-$23.5\%. The polynomial
parametrization of the background is indicated by the dashed line.
Red vertical lines represent invariant mass cuts. }
\end{center}
\label{fig2}
\end{figure}
The measurement in the centrality range $\sigma$/$\sigma_{TOT}$ =
5$-$23.5\% (called mid-central) is obtained by averaging the
results of bins 2 plus 3 with weights corresponding to the
fractions of the total cross section in these bins.

\section{Selection of particles}
The $\Lambda$ hyperon candidates were selected from the sample of
V$^0$-track configurations consisting of oppositely charged
particles, which include the $\Lambda$ decays into proton and
$\pi^{-}$ (branching ratio 63.9\%). Geometric and quality criteria
ensured that only reliably reconstructed tracks were processed.
The identification method \cite{Appelshauser:1999gs} relies on the
evaluation of the invariant mass distribution and is enhanced by
daughter particle identification applying a cut in $dE$/$dx$~
around the expectation value derived from a Bethe-Bloch
parametrization. $K^{0}_{s}$ mesons, which decay into $\pi^{+}$
and $\pi^{-}$ (branching ratio 68.6\%), were identified in a
similar way as $\Lambda$ hyperons. Instead of identification of
daughter particles by $dE$/$dx$ measurement we assigned the proton
mass to one of the daughter particles and excluded those pair
candidates which give entries near the expected $\Lambda$,
$\bar{\Lambda}$ invariant mass peak, 1.100 < $m_{p\pi^{-}}$,
$m_{\bar{p}\pi^{+}} < 1.132$ \gevcc~ (compare Fig.~3 (left) and
(right)). The yields of $\Lambda$ hyperons and $K^{0}_{s}$
particles are obtained by counting the number of entries in the
invariant mass peak above the estimated background as a function
of the azimuthal angle $\philab$ with respect to the event plane
(see below). The background is estimated from a fit of sum of
Lorentz distribution for $K^{0}_{s}$ or $\Lambda$ and a polynomial
background (Fig.~2 and Fig.~3). The number of accepted particles
in mid-central events is about 740.00 for $\Lambda$ hyperons in
the invariant mass window 1.108$-$1.124 \gevcc~and 440.000 for
$K^{0}_{s}$ mesons in the invariant mass window 0.488$-$0.508
\gevcc~. The acceptance of $\Lambda$ hyperons covers the range
$0.4 \lesssim \pt \lesssim 4.0$~\gevc~(see Fig.~4(left)) and $-1.5
\lesssim y \lesssim 1.0$ and strongly depends on $\pt$ and $y$.
The range of transverse momentum of $K^{0}_{s}$ mesons is more
narrow $0.2 \lesssim \pt \lesssim 3.5$~\gevc~(see Fig.~4(right)).
Multiplicative factors were introduced for every $\Lambda$
particle to correct the $\Lambda$ yields for detector and
reconstruction efficiency which depends on $\pt$ and $y$. Such an
additional correction has not been introduced  for $K^{0}_{s}$
mesons so far. $K^{0}_{s}$ elliptic flow data are still very
preliminary.
\begin{figure}
\begin{center}
\includegraphics[width=0.8\textwidth]{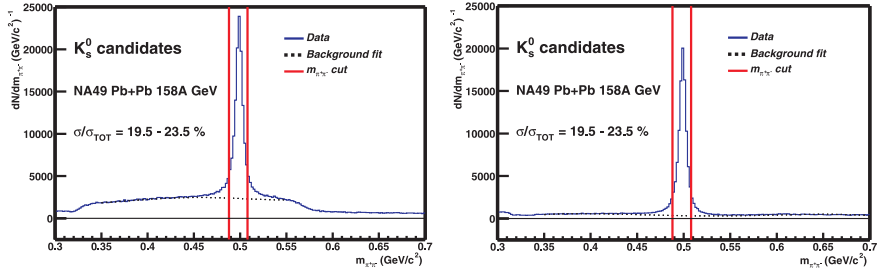}
\caption{The invariant mass distributions of $\pi^{+}\pi^{-}$
candidate pairs in Pb+Pb collisions at $\sqrt{s_{NN}}$ = 17.2 GeV
in the centrality range $\sigma$/$\sigma_{TOT}$ = 19.5$-$23.5\%.
The distribution is with (left) and without (right) admixture of
$p\pi^{-}$, $\bar{p}\pi^{+}$ candidate pairs with an invariant
mass near to the expected $\Lambda$, $\bar{\Lambda}$ hyperon peak
(see text for details). The polynomial parametrization of the
background is indicated by the dashed line. Red vertical lines
represent invariant mass cuts.}
\end{center}
\label{fig3}
\end{figure}

\begin{figure}
\begin{center}
\includegraphics[width=0.8\textwidth]{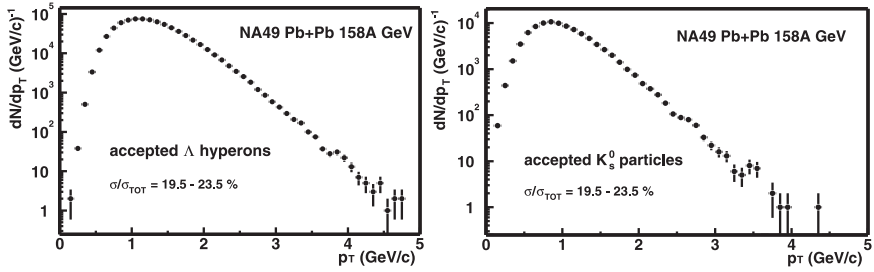}
\caption{Transverse momentum distributions for $\Lambda$ hyperons
(left) and $K^{0}_{s}$ particles (right) measured in the
centrality range $\sigma$/$\sigma_{TOT}$ = 19.5$-$23.5\% of Pb+Pb
collisions at $\sqrt{s_{NN}}$ = 17.2 GeV.}
\end{center}
\label{fig4}
\end{figure}

\section{Method}
The elliptic flow analysis is based on the standard procedure
outlined in \cite{Alt:2003ab,Voloshin:1999gs} to reconstruct the
event plane for each event and the corrections for the event plane
resolution. The event plane is an experimental estimator of the
true reaction plane and is calculated from the azimuthal
distribution of primary charged $\pi$ mesons. Identification of
pions is based on energy loss measurements ($dE$/$dx$) in the
TPCs. To avoid possible auto-correlations, tracks associated with
$\Lambda$ or $K^{0}_{s}$ candidates are excluded from the event
plane calculation. The method to determine the event plane angle
$\phiep$ uses the elliptic flow of pions, according to the
formula:
\begin{eqnarray}
& X_{2}&= \sum_{i=1}^{N}p^{i}_{\rb{T}}
[cos(2\phi^{i}_{\rb{lab}})-\langle
cos(2\philab)\rangle], \nonumber \\
& Y_{2}&= \sum_{i=1}^{N}p^{i}_{\rb{T}}
[sin(2\phi^{i}_{\rb{lab}})-\langle
sin(2\philab)\rangle],  \\
& \phiep&= tan^{-1}\left(\frac{Y_{\rb{2}}}{X_{\rb{2}}}\right)/2.
\nonumber
\end{eqnarray}
where $X_{\rb{2}},Y_{\rb{2}}$ are the components of the event
plane flow vector $\bf{Q_{\rb{{2}}}}$ and the sums run over
accepted charged pion tracks. The acceptance correction is based
on the recentering method of \cite{Alt:2003ab} which consists of
subtracting in Eq.~4.1 the mean values $\langle
cos(2\philab)\rangle$ and $\langle sin(2\philab)\rangle$. These
mean values are calculated in bins of $\pt$ and rapidity for all
charged pions in those events which contain at least one $\Lambda$
hyperon or $K^{0}_{s}$ meson candidate. The means were stored in a
3-dimensional matrix of 20 $\pt$ intervals, 50 rapidity intervals,
and eight centrality bins.

\begin{figure}
\begin{center}
\includegraphics[width=0.7\textwidth]{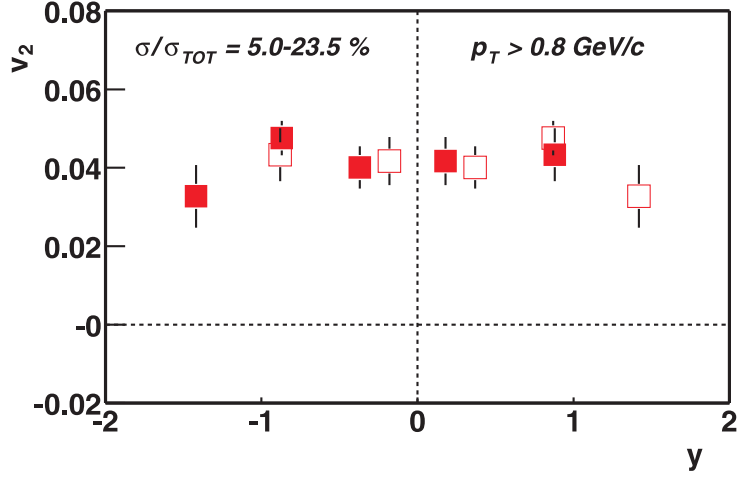}
\caption{Elliptic flow of $\Lambda$ hyperons measured in
mid-central Pb+Pb collisions as a function of rapidity. The open
points have been reflected about midrapidity. }
\end{center}
\label{fig5}
\end{figure}

\begin{figure}
\begin{center}
\includegraphics[width=0.8\textwidth]{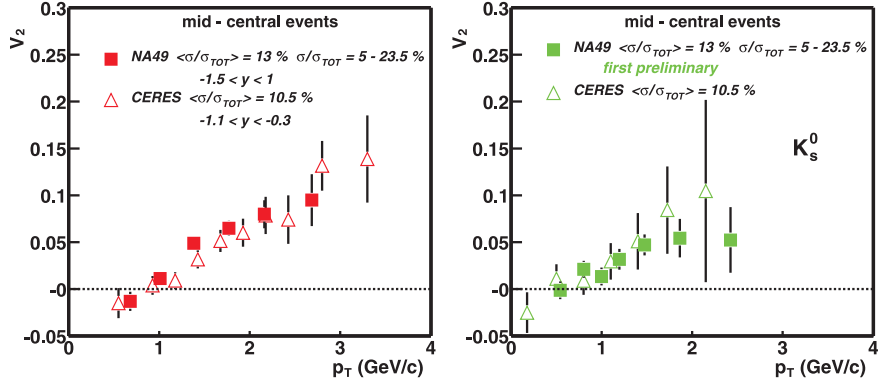}
\caption{Elliptic flow of $\Lambda$ hyperons (left) and
$K^{0}_{s}$ mesons measured by NA49 in Pb+Pb and NA45/CERES in
Pb+Au mid-central collisions as a function of transverse
momentum.}
\end{center}
\label{fig6}
\end{figure}

A second level acceptance correction is done by using mixed
events. We used 10 mixed events for each real event. Particles for
mixed events are randomly selected from different events in the
same centrality bin with at least one $\Lambda$ or $K^{0}_{s}$
particle. The final angular distributions were obtained by
dividing the real $\Lambda$ and $K^{0}_{s}$ angular distributions
by the mixed event distributions to remove the acceptance
correlations remaining after recentering. The corrected particle
azimuthal distributions are then fitted with a truncated Fourier
series:
\begin{eqnarray}
\frac{dN}{d(\philab-\phiep)} = \nonumber \\
   \mbox{const}\times(1 &+& v^{obs}_{2}cos[2(\philab-\phiep)]~~  \\
                 &+& v^{obs}_{4}cos[4(\philab-\phiep)]).~~ \nonumber
\end{eqnarray}
 The elliptic flow $\vi$ is
evaluated by dividing the observed anisotropy $v^{obs}_{\rb{2}}$
by the event plane resolution $R$:
\begin{equation}
v_{\rb{2}}=\frac{v^{obs}_{\rb{2}}}{R}.
\end{equation}
The event plane resolution,
\begin{equation}
R=\langle cos[2(\phiep-\phir)] \rangle=\sqrt{2\langle
cos[2(\Phi^{a}_{\rb{2EP}}-\Phi^{b}_{\rb{2EP}})]\rangle}, \nonumber
\end{equation}
is calculated from the correlation of two planes
($\Phi^{a}_{\rb{2EP}},\Phi^{b}_{\rb{2EP}}$) for random sub-events
with equal multiplicity. The results are $R$ = 0.27, 0.34 and 0.40
for centrality bins 1, 2 and 3, respectively. The uncertainty due
to the background subtraction, event plane resolution and the
mixed event corrections are added to the total error. The observed
hexadecupole anisotropy $v^{obs}_{\rb{4}}$ is consistent with zero
within statistical errors.

\begin{figure}
\begin{center}
\includegraphics[width=0.8\textwidth]{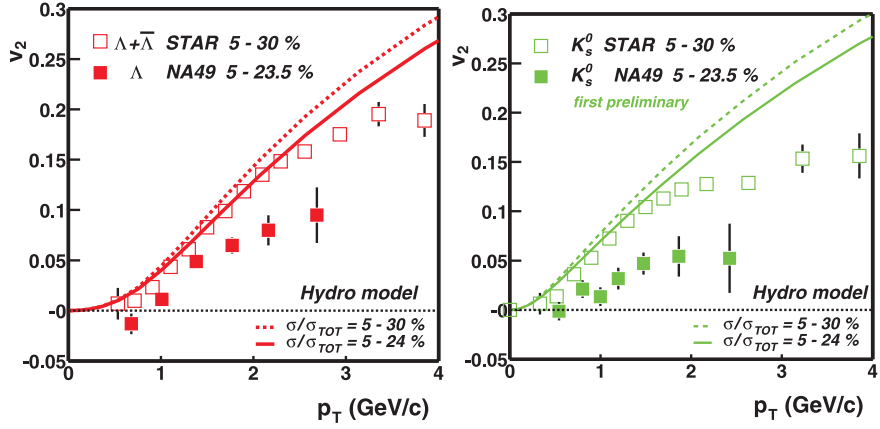}
\caption{Dependence of $v_{2}$ on $\pt$ for $\Lambda$ hyperons
(left) and $K^{0}_{s}$ mesons (right) measured at $\sqrt{s_{NN}}$
= 17.2 GeV and $\sqrt{s_{NN}}$ = 200 GeV in comparison to a
hydrodynamical calculation at $\sqrt{s_{NN}}$ = 200 GeV for two
centrality ranges.}
\end{center}
\label{fig7}
\end{figure}

\begin{figure}
\begin{center}
\includegraphics[width=0.7\textwidth]{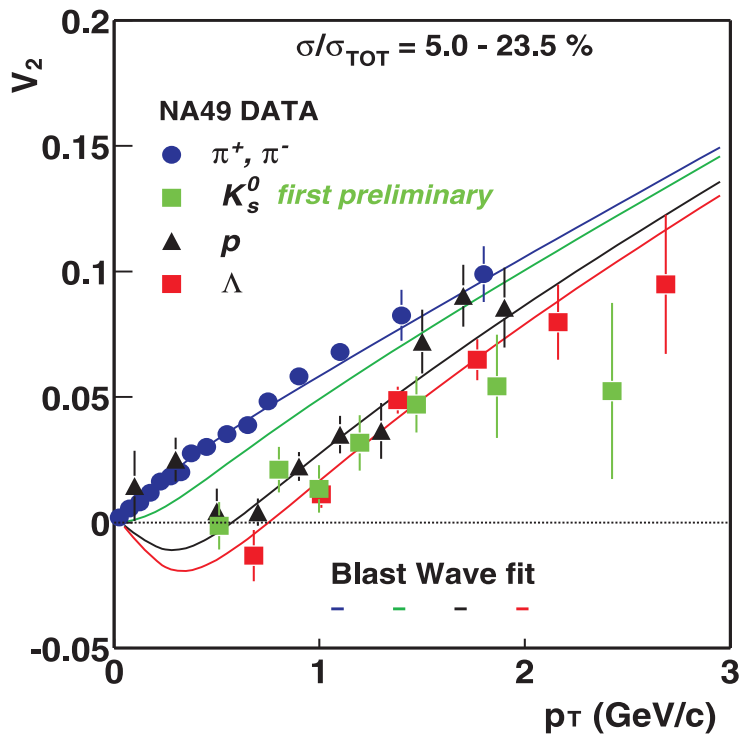}
\caption{Plot of $v_{2}$ of charged pions, protons, $\Lambda$
hyperons and $K^{0}_{s}$ mesons as a function of $p_{T}$ in
mid-central Pb+Pb collisions at $\sqrt{s_{NN}}$ = 17.2 GeV. The
solid lines represent results of a blast wave fit.}
\end{center}
\label{fig8}
\end{figure}

\section{Results}
Fig.~5 shows the $\pt$ averaged elliptic flow obtained from all
identified $\Lambda$ hyperons without $\pt$ cuts. It exhibits no
significant dependence on rapidity as was also observed for
protons (see Fig.~6 of \cite{Alt:2003ab}) in mid-central events.
The full rapidity range of the data in Fig.~5 was used to study
\vi~ as a function of $\pt$. The $\pt$ dependence of
rapidity-averaged $\Lambda$ and $K^{0}_{s}$ elliptic flow is shown
in Fig.~6(left) and Fig.~6(right) respectively in comparison to
CERES data\cite{CERES_QM05}.  The NA49 $\vi$ parameter
significantly increases with transverse momentum and agrees with
CERES results in Pb+Au mid-central collisions at the top SPS
energy. Fig.~7 shows a comparison of $\vi(\pt)$ of $\Lambda$
hyperons (left panel) and $K^{0}_{s}$ mesons (right panel) for
mid-central events measured by the NA49 and STAR experiments
\cite{Adams:2003am}. At SPS energy the elliptic flow grows
linearly with $\pt$ up to $\sim$2 GeV/c, but the increase is more
pronounced at RHIC than at SPS energy. It should be noted that
RHIC mid-central data have been measured in the centrality range
$\sigma/\sigma_{\rb{TOT}}$ = 5$-$30\% while SPS events are
somewhat more central. The effect of different centrality ranges
has been estimated by hydrodynamic calculations
\cite{Huovinen:2005gy} at RHIC energy for the slightly different
centrality bins of NA49 and STAR. As shown by the corresponding
curves in Fig.~7 this explains only partly the difference between
both measurements. A comparison of $\vi(\pt)$ for pions, protons,
$\Lambda$ hyperons and $K^{0}_{s}$ mesons as measured by the NA49
experiment in mid-central events is displayed in Fig.~8. The
values for pions and protons were obtained as the cross section
weighted averages of the measurements published in
\cite{Alt:2003ab} for the appropriate centrality range. As seen in
Fig.~8 the elliptic flow grows linearly with $\pt$ for all
particle species but the rise for pions starts from $\pt$ close to
zero while for protons, $\Lambda$ and $K^{0}_{s}$ mesons it starts
from $\pt\approx0.5$~\gevc. The elliptic flow for pions is
significantly larger than that for heavier particles although at
$\pt \approx 2$~\gevc~the flow becomes similar for all particle
species. Except for $K^{0}_{s}$ mesons the measurements are
reproduced by blast wave fits
\cite{Huovinen:2001cy,Adler:2001nb,Retiere_Lisa} (curves in
Fig.~8) with the following parameters: freeze-out temperature
$T$~=~95 MeV, mean transverse expansion rapidity
$\rho_{0}$~=~0.85, its second harmonic azimuthal modulation
amplitude $\rho_{2}$~=~0.021 and the variation in the azimuthal
density of the source elements $s_{2}$~=~0.035. In Fig.~9(left)
the measured values of $\vi$ are compared to hydrodynamical model
calculations \cite{Huovinen_private} assuming a first-order phase
transition to a QGP at the critical temperature $T_{\rb{c}}=165$
MeV. With the freeze-out temperature $T_{\rb{f}}=120$ MeV tuned to
reproduce particle spectra, the model calculations significantly
overestimate the SPS results for semi-central collisions (full
curves in Fig.~9(left)) in contradiction to predictions at RHIC
energy which agree with data quite well for $\pt \lesssim
2$~\gevc~\cite{Adams:2003am}. The discrepancy at SPS may indicate
a lack of complete thermalisation or a viscosity effect. On the
other hand, the model reproduces qualitatively the characteristic
hadron-mass ordering of elliptic flow. Thus the data support the
hypothesis of early development of collectivity. The calculation
from the same model with a higher temperature $T_{\rb{f}}=160$ MeV
exhibits better agreement with the $\Lambda$ flow data (dotted
curves in Fig.~9(left)). Unfortunately, the model does not
simultaneously reproduce the $\mt$ spectra with such a high
freeze-out temperature. A partial solution of this problem can be
found by coupling a hadronic rescattering phase to the
hydrodynamical evolution and hadronisation \cite{Teaney:2000cw}.

\begin{figure}
\begin{center}
\includegraphics[width=0.8\textwidth]{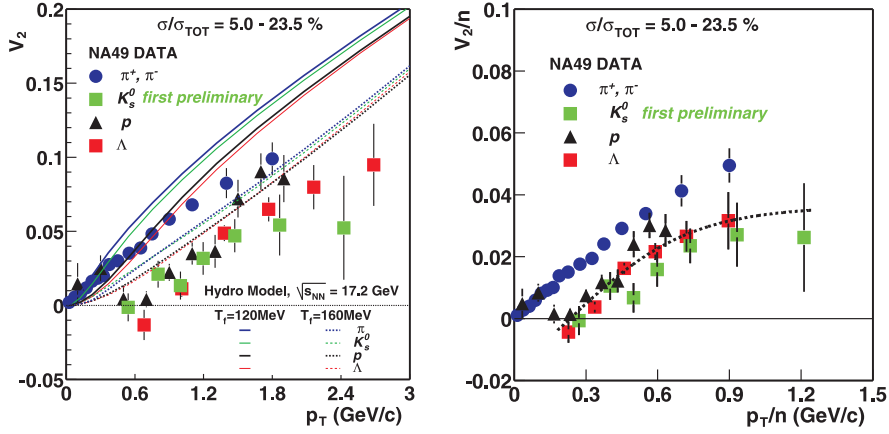}
\caption{Left: $v_{2}$($p_{T}$) of charged pions, protons,
$\Lambda$ hyperons and $K^{0}_{s}$ mesons from 158A GeV Pb+Pb
mid-central collisions. Hydrodynamic model calculations at
$\sqrt{s_{NN}}$ = 17.2 GeV are shown as solid ($T_{f}$=120 MeV)
and dashed ($T_{f}$=160 MeV) lines (see text for details). Right:
$v_{2}$ scaled by the number of quark as a function of scaled
$p_{T}$. All data are from 158A GeV Pb+Pb mid-central collisions.
The dashed line is the scaled result of the fit to p, $\Lambda$
and $K^{0}_{s}$. }

\end{center}
\label{fig9}
\end{figure}

 Coalescence models predict that \vi~
will approximately scale with the number $n$ of constituent
quarks. When plotting at the similarly scaled value $\pt/n$
results for all hadrons are expected to fall on an universal. This
prediction agrees well with RHIC data in the intermediate
transverse momentum region $\pt/n$
> 0.7 \gevc~\cite{Adams:2003am}. The universal curve for all
hadrons is expected to represent the momentum-space anisotropy of
constituent quarks prior to hadron formation. The naive
coalescence model roughly agrees with our proton, $\Lambda$ and
$K^{0}_{s}$ at higher $\pt$ values ( Fig.9(right)) although the
$\pt$ range of accepted particles is too narrow and error bars are
to large to draw a clear conclusion. In more realistic quark
coalescence model the resonance decays and quark momentum
distribution in hadrons can lead to higher \vi~ for pions and
generally deviations of meson elliptic flow from the scaling
behavior as observed on Fig.~9(right) and also seen by the STAR
experiment.

\begin{acknowledgments}
This work was supported by the US Department of Energy Grant
DE-FG03-97ER41020/A000, the Bundesministerium fur Bildung und
Forschung 06F-137, Germany, the Virtual Institute VI-146 of
Helmholtz Gemeinschaft, Germany, the Polish State Committee for
Scientific Research (1 P03B 006 30, 1 P03B 097 29, 1 P03B 121 29,
1 P03B 127 30), the Hungarian Scientific Research Foundation
(T032648, T032293, T043514), the Hungarian National Science
Foundation, OTKA, (F034707), the Polish-German Foundation, the
Korea Science \& Engineering Foundation (R01-2005-000-10334-0) and
the Bulgarian National Science Fund (Ph-09/05).
\end{acknowledgments}

\end{document}